\documentclass[fleqn,usenatbib]{mnras}

\usepackage{newtxtext,newtxmath}
\usepackage[T1]{fontenc}
\usepackage{ae,aecompl}

\usepackage{graphicx}	% Including figure files
\usepackage{amsmath}	% Advanced maths commands
\usepackage{amssymb}	% Extra maths symbols

\title[WD1032$+$011 an eclipsing WD-BD binary]{WD1032$+$011, an inflated brown dwarf in an old eclipsing binary with a white dwarf}
\author[S. L. Casewell et al]{S. L. Casewell$^{1, \dagger}$ \thanks{E-mail: slc25@le.ac.uk}, C. Belardi$^{1}$,  S. G. Parsons$^{2, \dagger}$, S. P. Littlefair$^{2}$, I.P. Braker$^{1}$ \newauthor J. J. Hermes$^{3}$, J. Debes$^{4}$, Z. Vanderbosch$^{5,6}$,  M.R. Burleigh$^{1}$, B. T. G\"ansicke$^{7}$, \newauthor V.~S. Dhillon$^{2,8}$, T.~R. Marsh$^{7}$, D.~E. Winget$^{5,6}$, K.~I. Winget$^{5,6}$\\
$^{\dagger}$STFC Ernest Rutherford Fellow \\
$^{1}$School of Physics and Astronomy, University of Leicester, University Road, Leicester LE1 7RH, UK \\
$^{2}$Department of Physics and Astronomy, University of Sheffield, Sheffield, S3 7RH, UK\\
$^{3}$Department of Astronomy, Boston University, Boston, MA-02215, USA\\
$^{4}$Space Telescope Science Institute, Baltimore, MD 21218, USA\\
$^{5}$Department of Astronomy, University of Texas at Austin, Austin, TX, 78712, USA\\
$^{6}$McDonald Observatory, Fort Davis, TX-79734, USA\\
$^{7}$Department of Physics, University of Warwick, Gibbet Hill Road, Coventry CV4 7AL, UK\\
$^{8}$Instituto de Astrof\'isica de Canarias (IAC), E-38200 La Laguna, Tenerife, Spain}

\date{Accepted XXX. Received YYY; in original form ZZZ}

\pubyear{2020}

\begin{document}
\label{firstpage}
\pagerange{\pageref{firstpage}--\pageref{lastpage}}
\maketitle

\begin{abstract}
We present the discovery of only the third brown dwarf known to  eclipse a non-accreting white dwarf. $Gaia$ parallax information and multi-colour photometry confirm that the white dwarf is cool (9950$\pm$150~K) and has a low mass (0.45$\pm$0.05~M$_{\odot}$), and spectra and lightcurves suggest the brown dwarf has a mass of 0.067 $\pm$0.006 M$_{\odot}$ (70 M$_{\rm Jup}$) and a spectral type of L5$\pm$1. The kinematics of the system show that the binary is likely to be a member of the thick disk and therefore at least 5~Gyr old. The high cadence lightcurves show that the brown dwarf is inflated, making it the first brown dwarf in an eclipsing white dwarf-brown dwarf binary to be so.
\end{abstract}

\begin{keywords}
brown dwarfs, eclipsing binaries, white dwarfs
\end{keywords}

%%%%%%%%%%%%%%%%%%%%%%%%%%%%%%%%%%%%%%%%%%%%%%%%%%

%%%%%%%%%%%%%%%%% BODY OF PAPER %%%%%%%%%%%%%%%%%%

\section{Introduction}

Since 1995 there have been many brown dwarfs discovered, however the majority of these are isolated field objects. Indeed, in the search for "benchmark" objects, only two double-lined eclipsing brown dwarf systems have been discovered, 2MASS~J05352184$-$0546085 \citep{stassun} and 2MASSW~J1510478-281817 \citep{triaud20}. Both of these systems are young, with ages $<$ 100~Myr meaning they are not good tests of evolutionary models for field objects. Recently studies of open star clusters with K2 and CoRoT have discovered more young eclipsing brown dwarf binaries (e,g, \citealt{david19, nowak}), however there are still none known in the field.

Indeed, despite the success of K2 and SuperWASP at discovering hot Jupiter exoplanets, there are very few brown dwarfs that have been discovered in similar eclipsing systems. Only $\sim$20 are known in close orbits around main sequence stars (e.g. \citealt{carmichael20} and references therein), and only six have been confirmed in their evolved form in post-common envelope binaries. Four of these brown dwarfs orbit hot subdwarfs (sdB: \citealt{geier11, schaffenroth14, schaffenroth15}) and two orbit white dwarfs, SDSS~J141126.20$+$200911.1 \citep{beuermann13, littlefair14} and SDSS ~J120515.804$-$024222.6 \citep{parsons17}.  Often, the brown dwarf atmospheres in these systems are affected by the intense irradiation from their host star, resulting in large atmospheric differences between the day and night sides \citep{beatty} which can include emission lines from a chromosphere on the brown dwarf (e.g. \citealt{longstaff17}), or signs of photochemistry \citep{casewell18b, casewell15}. In many of these brown dwarfs orbiting main sequence stars, the brown dwarf is also inflated, meaning these binaries are unsuitable to use as benchmark systems to test the brown dwarf mass-radius relation.  One way of searching for suitable benchmark systems is to use eclipsing brown dwarfs orbiting cool white dwarfs (T$_{\rm eff}<$10,000~K)  where there is only a small amount of irradiation impacting the brown dwarf atmosphere, and consequently, no emission is seen.

We present here the discovery of a new eclipsing detached post-common envelope system, WD1032+011AB, Hereafter WD1032+011. WD1032+011 was first identified as a hydrogen rich white dwarf (DA) by \citet{vennes02} in the 2df QSO survey. \citet{eisenstein06} used SDSS spectra to measure an effective temperature of 9904$\pm$109 K and log $g$ of 8.13$\pm$0.15. \citet{steele11} suggested WD1032+011 had an infrared excess suggestive of an L5 companion with a mass of 55$\pm$4 M$_{\rm Jup}$ based on its UKIDSS and SDSS magnitudes. We have used optical and near-infrared (NIR) spectroscopy and optical lightcurves to confirm there is indeed a brown dwarf secondary in the system, and that it totally eclipses the white dwarf. This discovery increases the number of eclipsing white dwarf-brown dwarf binaries to three, and provides constraints on how the radius of a brown dwarf is affected by irradiation and the common-envelope phase.

\section{Observations}
\begin{table}
	\centering
	\caption{Position and magnitudes for WD1032+011.}
	\label{tab:info}
	\begin{tabular}{ccc} % 3 columns, alignment for each
    \hline
	Property	&	Value		&Survey\\
	\hline
	Ra (J2000)&10:34:48.93 &\\
    Dec (J2000)&+00:52:01.4&\\
    Distance (pc)&326.78$\pm$36.92&Gaia DR2\\
 %   M$_{\rm wd}$& 0.68$\pm$0.1 M$_{\odot}$& \citet{steele11}\\
    fuv&22.670$\pm$0.224&\textit{Galex}\\
    nuv&20.152$\pm$0.037&\textit{Galex}\\
    $u$&19.547$\pm$0.032&SDSS\\
    $g$&19.034$\pm$0.010&SDSS \\
    $r$&19.076$\pm$0.012&SDSS \\
    $i$&19.169$\pm$0.019&SDSS \\
    $z$&19.240$\pm$0.076&SDSS \\
    $Y$&18.820$\pm$0.042 &UKIDSS LAS \\
    $J$&18.648$\pm$0.056  &UKIDSS LAS \\
    $H$&18.202$\pm$0.107& UKIDSS LAS\\
    $K$&18.034$\pm$0.141&UKIDSS LAS\\
\hline
\end{tabular}
\end{table}

\subsection{Kepler photometry}

\begin{figure*}
 \begin{center}
 \includegraphics[scale=0.7]{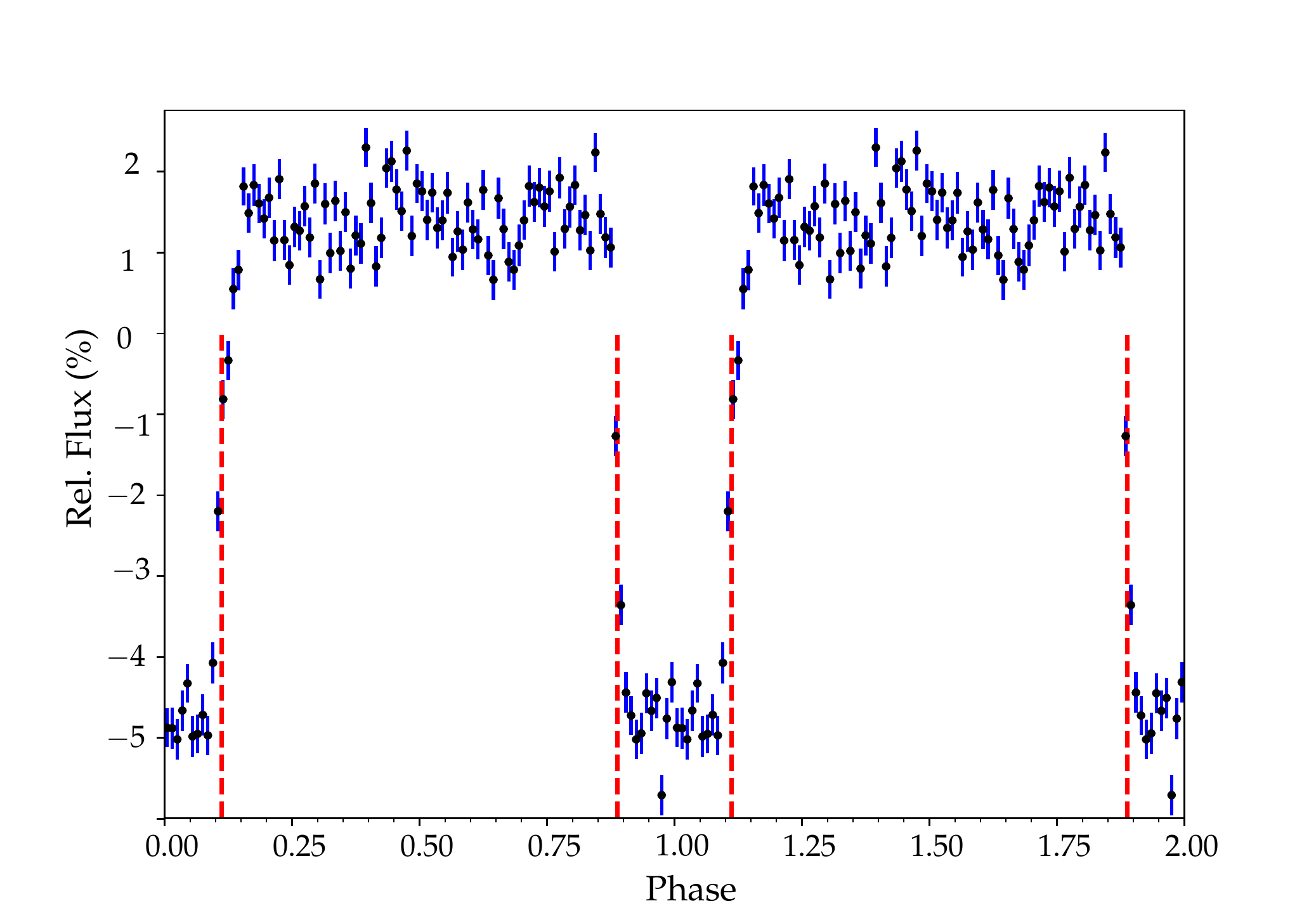}
    \caption{The phase folded and binned (into 100 bins) K2 light curve of WD1032+011 on a period of 0.092 days. The primary eclipse is clearly evident, and the red lines represent the 29.4 min cadence of the K2 data, showing the undersampling of the eclipse. The lightcurves have been duplicated over two periods for display purposes.}
     \label{fig:k2binnedlc}
     \end{center}
 \end{figure*}

The Kepler 2 (K2) mission \citep{howell2014} has observed over 1500 spectroscopically and photometrically identified white dwarfs up to Campaign 15. WD1032+011 was proposed as a K2 target (EPIC248433650) in three separate proposals (PI Burleigh, PI Hermes and PI Redfield) in the campaign 14 field (centered on J2000, RA 10:42:44, dec 06:51:06). The Kepler 2 data release for campaign 14 included the calibrated pixel files and a standard pipeline lightcurve by the Kepler Guest Observer Office \citep{vancleve2016}. See Table \ref{tab:info} for the photometric parameters. WD1032+011 was observed by K2 for $\approx81$ days between 2017 May 31 and 2017 Aug 19 in long cadence mode. Our analysis used the K2 pixel file downloaded from the Mikulski Archive for Space Telescopes (MAST). Due to significant cross talk on this area of the Kepler CCD a custom mask was used to create a lightcurve. The lightcurve was normalised and flagged points (such as those affected by cosmic ray hits) were removed, resulting in a lightcurve with 3233 data points. We searched for periodicity in the lightcurve using the Lomb-Scargle \citep{Lomb1976,Scargle1982} routine in \textsc{idl} and the periodogram software packages \textsc{Vartools} \citep{hartman2016}. These identified a most likely period of $0.09155900043(3)$ days ($\approx$2.2 hours) with period uncertainties determined using the bootstrap resampling (with replacement) methodology detailed in \citealt{lawrie13_1}. The binned and phase-folded lightcurve (Fig. \ref{fig:k2binnedlc}) clearly shows a primary eclipse. Due to the long cadence (29.4 minutes) of the K2 observations the depth is shallower and broader than would be expected for a companion to the white dwarf at this period.  There is no evidence of a secondary eclipse in the K2 lightcurve.

\subsection{FORS spectroscopy}

\label{sec:spec} % used for referring to this section from elsewhere
We observed WD1032$+$011 with the visual and near-UV FOcal Reducer and low-dispersion Spectrograph (FORS; \citealt{fors}) on the Very Large Telescope in service mode as part of programme 098.D-0717(A). We used the G1200B grism and the 1" slit to cover the majority of the Balmer lines including H$\beta$ at a resolution $\lambda/\Delta\lambda$$\sim$ 1400.  We obtained an hour of data on each of the nights of 2016 Dec 05, 2016 Dec 06 and 2016 Dec 24. Each hour was split into six exposures of 420~s, providing 18 spectra in total. 

The data were reduced using the FORS specific \textsc{reflex} \citep{reflex} pipeline.  The spectra were then normalised in \textsc{molly} and phase folded according to the ephemeris determined from the K2 data. The spectra clearly show that the white dwarf is moving on the orbital period determined from the K2 data, however they are not of sufficient resolution to determine the systemic velocity, $\gamma$, or the radial velocity, $K_1$, of the white dwarf and so were not included in the subsequent fitting presented here.

 \subsection{GMOS spectroscopy}
 As our FORS spectra were not of sufficient resolution to measure the radial velocity of the white dwarf, we observed WD1032$+$011 on the nights of 2019 Jan 11, 2019 Jan 12 and 2019 Jan 13 with the Gemini Multi-Object Spectrograph (GMOS: \citealt{hook}) on Gemini-North in service mode as part of programme GN-2019A-Q-227 (PI: Debes). We obtained 24 spectra with the R831 grating with a central wavelength of 5750~ \AA, resulting in a wavelength range of 4600-6900~\AA. We used the 0.75" slit, 2$\times$2 binning and 900~s exposures to obtain resolution of ($\lambda/\Delta\lambda$) $\sim$ 4000. The airmass was between 1" and 1.5" for the observations. 
 
 The data were reduced using the GMOS specific packages in \textsc{iraf} \citep{tody} for long-slit spectra before being calibrated using a standard star observation of Hiltner 600.
 
 \subsection{GNIRS spectroscopy}
We observed WD1032-011 with Gemini North and the cross-dispersed spectrograph GNIRS \citep{gnirs} as part of programme GN-2019A-Q-227 (PI: Debes). We used the short camera with the 1.00" slit providing a resolution of ($\lambda/\Delta\lambda$)$\sim$500 over the whole 0.8-2.5 micron spectrum. We nodded the observations with 440 s exposures taken at each nod point and combined the 8 exposures at the reduction stage. 
The data were reduced using \textsc{spextool} v4.1 \citep{cushing04} which had been adapted for use with GNIRS (K. Allers, private comm.) and telluric corrected using \textsc{xtellcorr} \citep{vacca03} and an A0V standard star.

 \subsection{McDonald 2.1-m Photometry}

We acquired high-speed time-series photometry of WD1032$+$011 on four consecutive nights, 2017 Dec 18--21, using the Princeton Instruments ProEM frame-transfer CCD on the McDonald Observatory 2.1-m Otto Struve telescope. The nights were clear, with a new moon, but poor seeing. Each night we used an Astrodon Gen2 Sloan $g'$ filter, and in total our observations covered six eclipses of the white dwarf primary. See Table~\ref{tab:mcd} for a summary of observing information and conditions. 

Using standard calibration frames taken before each night of observation, we bias, dark, and flat-field corrected the McDonald images with {\sc iraf}. We performed circular aperture photometry using the {\sc iraf} routine {\sc ccd\_hsp} \citep{Kanaan2002}. Background counts were subtracted using an annulus placed around each aperture. We performed the aperture photometry both forward through each ingress and backward through each egress to ensure the aperture was properly placed on the centroid of our target as it went in and out of total eclipse.

To generate light curves for each night, we used the {\sc Wqed} software suite \citep{Thompson2013}. We divided the target photometry by two nearby comparison stars, the same for each night, and then normalised by the target's mean intensity. We clipped any outliers or heavily cloud-affected data from the light curves and then selected the optimal aperture size which minimised the average point-to-point scatter when out of eclipse. Lastly, we used {\sc Wqed} to apply a barycentric correction to the mid-exposure timestamp of each image.

\begin{table}
	\centering
	\caption{McDonald 2.1-m observations of WD1032$+$011.}
	\label{tab:mcd}
	\begin{tabular}{cccccc}
		\hline
		Night&Filter&Seeing&Airmass&Exposure&Duration\\
		&&($"$)&&(s)&(h)\\
		\hline
		2017 Dec 18&$g'$&$2.0$&1.18$-$1.15&10&0.66\\
		2017 Dec 19&$g'$&$2.3$&1.31$-$1.20&10&2.93\\
		2017 Dec 20&$g'$&$3.0$&1.46$-$1.22&15&3.87\\
		2017 Dec 21&$g'$&$3.6$&1.41$-$1.25&15&3.83\\
		\hline
	\end{tabular}
\end{table}

\subsection{ULTRACAM photometry}
We observed three eclipses of WD1032$+$011 using ULTRACAM \citep{dhillon} on the ESO New Technology Telescope. The observing information and conditions are listed in Table \ref{tab:ucam}. ULTRACAM observes in three filters simultaneously, with dead time of $\sim$25~ms, and we used on chip coadding in the $u'$ band to increase the exposure time to between 13.5 and 18~s, hence increasing the signal-to-noise ratio in this filter.

\begin{table}
	\centering
	\caption{NTT/ULTRACAM observations of WD1032$+$011.}
	\label{tab:ucam}
	\begin{tabular}{ccccc} % 3 columns, alignment for each
    \hline
	Night	&	Filters		&Seeing & Airmass & Exposure\\
	    &   & (")& & (s)\\
	\hline
2018 Jan 19&$u'g'r'$&$1.4$&1.24$-$1.36&6\\
2018 Jan 23&$u'g'r'$&$1.0$&1.56$-$1.77&6\\
2019 Mar 01&$u'g'i'$&$0.7$&1.41$-$1.49&4.5\\
\hline
\end{tabular}
\end{table}

The data were reduced using the ULTRACAM pipeline software.  The source flux was determined using aperture photometry with a variable aperture scaled according to the full width at half-maximum. Any variations in observing conditions were accounted for by determining the flux relative to a comparison star within the field of view. 

We performed a fit to each lightcurve using  \textsc{lcurve} \citep{copperwheat10} to determine $T0$, and the continuum flux level, as the photometry on each night was performed using a different reference star. %T

\section{Results}
\subsection{Ephemeris}
We used the initial ephemeris as determined from the K2 data, and fit each individual eclipse from the ULTRACAM and ProEM instruments in the $g$ band with the lightcurve fitting code  \textsc{lcurve} \citep{copperwheat10} to determine the centre of each eclipse. We selected an eclipse that lay in the middle of the observed times to represent cycle 0 (the first eclipse on the 2017 Dec 21), and determined the cycle of each eclipse using the K2 period, before fitting the function $T=T_0 + E*P$, where $T$ is time of eclipse, $E$ is the cycle number and $T_0$ and $P$ are the zeropoint and period respectively, to the data.  The linear ephemeris was fit using the affine-invariant MCMC sampler \textsc{emcee} \citep{foreman13}. The cycle numbers were adjusted to minimise the covariance between $T_0$ and $P$. The final ephemeris is given in Table \ref{tab:res}. The best fit has a  $\chi^2$ of 3.9 with 7 degrees of freedom. 
The individual eclipse times, including the uncertainties, are given in the Appendix (Table \ref{tab:eclipse}).

\subsection{Radial Velocity}
\begin{figure*}
 \begin{center}
 \includegraphics[scale=0.5, trim = 0cm 0cm 0cm 0cm, clip]{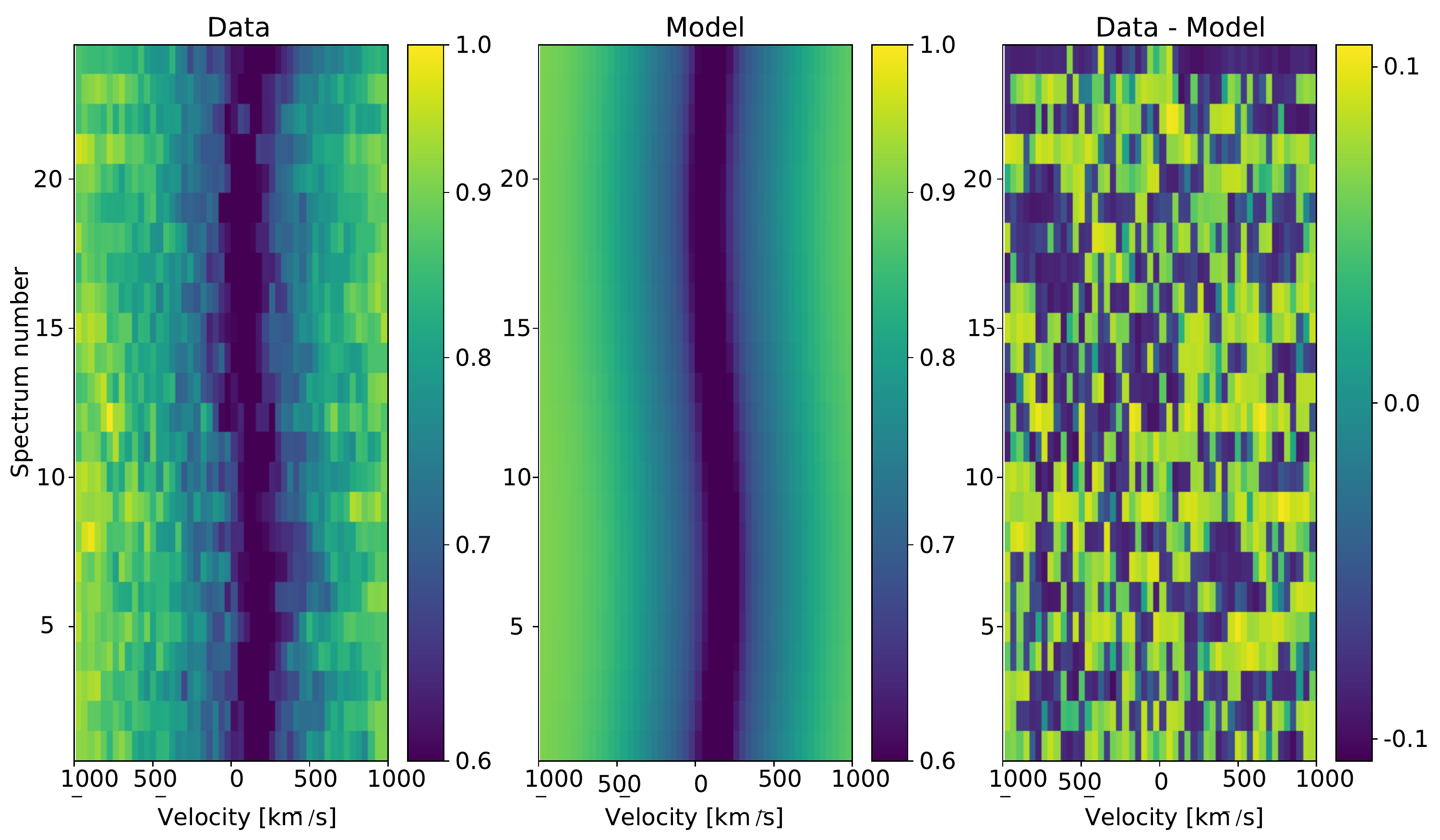}
    \caption{Trailed GMOS spectra showing one full orbit, centred on the H$\beta$ line shown with the data (left), model generated from the Gaussian fitting (centre), and the fit residuals (right). }
     \label{fig:trail}
     \end{center}
 \end{figure*}

The GMOS spectra were analysed using the \textsc{molly} software\footnote{http://deneb.astro.warwick.ac.uk/phsaap/software/molly/} package. We analysed the spectra using the techniques described by \citet{parsons17}, fitting the orbit using all of the spectra, not determining velocities for each individual spectrum. We normalised the spectra by dividing by the average spectrum for each hour-long observing block, phase folded the spectra using the ephemeris derived from the eclipses (in the previous subsection), and fitted the H$\beta$ absorption line using three Gaussians, two broad and one narrow. The three Gaussians were fixed to have the same $K$ and $\gamma$ velocities, and were allowed to vary around the orbit as $\gamma_1 + K_1 \sin(2\pi\phi)$, where $\phi$ is the orbital phase. The errors on the radial velocity parameters  were determined by adding 1 kms$^{-1}$ in quadrature in order to achieve a reduced chi-squared of $\chi_{\nu}^2\sim$~1. The final velocities were $K_1$=48.8$\pm$2.6~kms$^{-1}$ and $\gamma$=122.1$\pm$1.9~kms$^{-1}$. The trailed spectra and model are shown in Figure \ref{fig:trail}.

We also searched the spectra for any other emission or absorption lines that may have been in our wavelength range, notably Na I and Mg I, but none were present.

\subsection{Effective temperature and log $g$}
\begin{figure}
 \begin{center}
 \includegraphics[trim = 0cm 0cm 0cm 0cm, clip, width=\columnwidth]{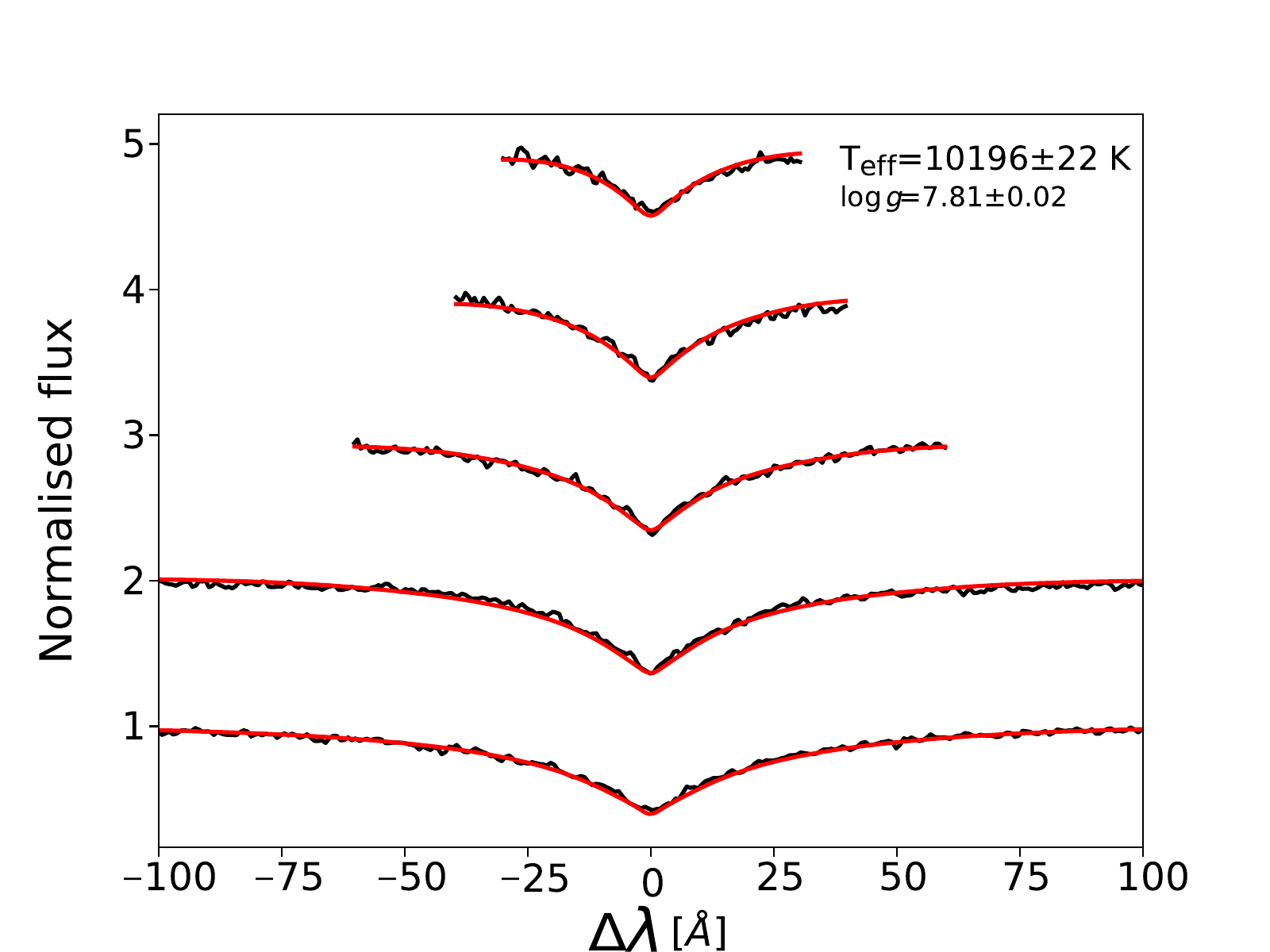}
    \caption{Combined FORS2 spectrum (black, H$\beta$ to H8, bottom to top) of WD1032+011 with the best fitting model $T_{eff}$ = 10196 $\pm$ 72~K and log $g$ = 7.81$\pm$0.02 overplotted in red. These errors are the fitting errors only. Please see the main text for an explanation of the likely true errors.}
     \label{fig:teff}
     \end{center}
 \end{figure}

We combined all 18 of the FORS spectra by shifting them into the rest frame of the white dwarf using the K$_1$ value from the radial velocity fitting, and then used DA white dwarf models from \citet{koester2010} to determine the effective temperature and gravity of the white dwarf. Our model grid consists of a set of DA white dwarf model spectra with mixing length ML2/$\alpha$ = 0.8 computed on a grid of 6.5 $\geq$ log $g$ $\geq$ 9.5 in steps of 0.25 dex and 6000$\geq$ $T_{eff}$ $\geq$ 40000 K in steps of 1000 K. Our parameters have errors that are underestimated as suggested by \citet{napiwotzki99} so we follow their method and assume an uncertainty of 2.3 per cent in T$_{\rm eff}$ and 0.07 dex in log g. The new parameters are $T_{ eff}$ = 10196 $\pm$ 235~K and log $g$ = 7.81$\pm$0.07.  (Figure \ref{fig:teff}). These values have a much lower log $g$ compared to the \citet{eisenstein06} values. However, they still predict a white dwarf far more luminous than the $Gaia$ parallax implies (such a white dwarf would be expected to have an absolute magnitude of $M_G=9.3$ compared to the actual value of $M_G=11.8$, \citep{Gaia18}). We decided to independently measure the white dwarf parameters by fitting the spectral energy distribution (SED) of WD1032+011 with white dwarf models, including the $Gaia$ parallax information within the fit.

We retrieved broadband photometry of WD1032+011 from GALEX, SDSS and UKIDSS (see Table~\ref{tab:info}) and fitted these with DA white dwarf spectra from \citet{koester2010}. For a given combination of log $g$ and $T_{eff}$ we used the mass-radius relation of \citet{panei07} for He core white dwarfs to estimate the white dwarf radius, which, combined with the parallax, was used to scale the model spectrum. We also included the effects of reddening. We discarded the $z$, $Y$, $J$, $H$ and $K$-band measurements since the brown dwarf contributes a non-negligible amount of flux at these wavelengths. Model parameters and their uncertainties were found using the Markov Chain Monte Carlo (MCMC) method \citep{Press07} implemented using the python package {\sc emcee} \citep{foreman13}, where the likelihood of accepting a model was based on a combination of the $\chi^2$ of the SED fit and prior probabilities on the parallax (Gaussian, based on the $Gaia$ measurement and associated uncertainty) and the reddening (uniform from zero up to the  maximum possible value of 0.052 based on reddening maps, \citealt{Schlafly11}). The final parameters are listed in Table~\ref{tab:res} and are consistent with the values from the fit to the FORS2 spectrum but with a slightly lower surface gravity (thus lower mass). The somewhat larger log $g$ implied from the spectral fit may be due to some minor contamination of the Balmer lines by emission from the brown dwarf, although there is no clear evidence of emission lines in the FORS2 spectra.

%T$_{\rm eff}$(K)&10196 $\pm$235& SED\\
%log  g&7.81$\pm$0.07 SED\\

%Using these values of effective temperature and log g with the thick-H layer evolutionary models of \citet{fontaine01} give a radius of 0.145 R$_{\odot}$, a mass of 0.496 M$_{\odot}$ and a cooling time of 0.455 Gyr. %Gravitational redshift is 21.8 kms$^{-1}$.

\begin{figure}
 \begin{center}
 \includegraphics[trim = 0cm 0cm 0cm 0cm, clip, width=\columnwidth]{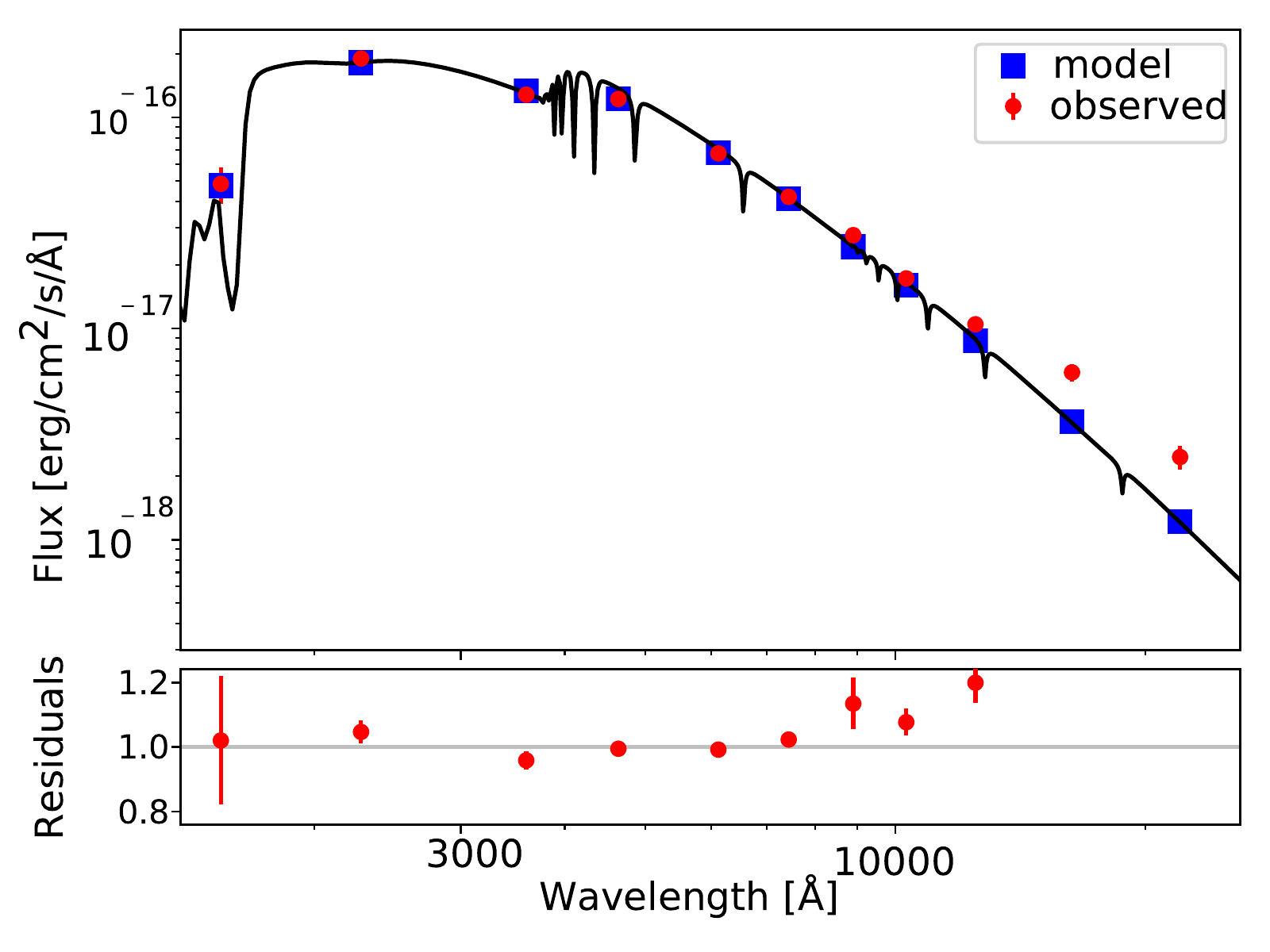}
    \caption{Spectral energy distribution of WD1032+011 (red points are GALEX, SDSS and UKIDSS measurements) with the best-fit model white dwarf spectrum (black line and blue points). Only measurements blueward of the $z$-band were included in the fit. The lower panel shows the residuals to the fit, which rise rapidly at longer wavelengths where the contribution from the brown dwarf is no longer negligible. The final parameters from this fit are $T_{eff}$=9950 $\pm$150~K and log  $g$=7.65$\pm$0.13.}
     \label{fig:sed}
     \end{center}
 \end{figure}

%T0 - MBJD(TDB)  58108.3988957
%P 0.091558799999999996D0 

%FINAL VALUES
%Total Mass = 0.516814 +/- 0.054896
%WD Mass = 0.450273 +/- 0.050047
%WD Rad  = 0.014759 +/- 0.001325
%BD Mass = 0.066541 +/- 0.006146
%BD Rad  = 0.105197 +/- 0.010108
%a       = 0.685421 +/- 0.024492
%parallax: 3.07 +/- 0.27 (Gaia parallax is 3.06 +/- 0.39)
%E(B-V) = 0.03 +/- 0.01 (I set an upper limit on this of 0.052 from dust maps)
%Teff = 9950 +/- 150 K
%log(g) = 7.65 +/- 0.13.
\subsection{Spectral type of the secondary}
To determine the spectral type of the brown dwarf secondary we created combined white dwarf-brown dwarf models. We used a white dwarf model from \citet{koester2010} using our derived $T_{eff}$ and log $g$ and the model absolute magnitudes of a lone white dwarf with these parameters from \citet{holberg06} and \citet{tremblay11} to create a normalised white dwarf model at 10 pc. We then repeated this process using the absolute magnitudes of brown dwarfs with spectral types L3, L4, L5 and L6 from \citet{dupuy12} and brown dwarf template spectra from \citet{cushing05} and \citet{rayner09} archived in the IRTF spectral library.
Ensuring both sets of spectra were on the same wavelength scale, we then combined them, and normalised the models to the SDSS $i$ band, where the brown dwarf contribution to the total flux is negligible, as can be seen from the SED fitting to the white dwarf photometry.

\begin{figure}
 \begin{center}
 \includegraphics[width=\columnwidth]{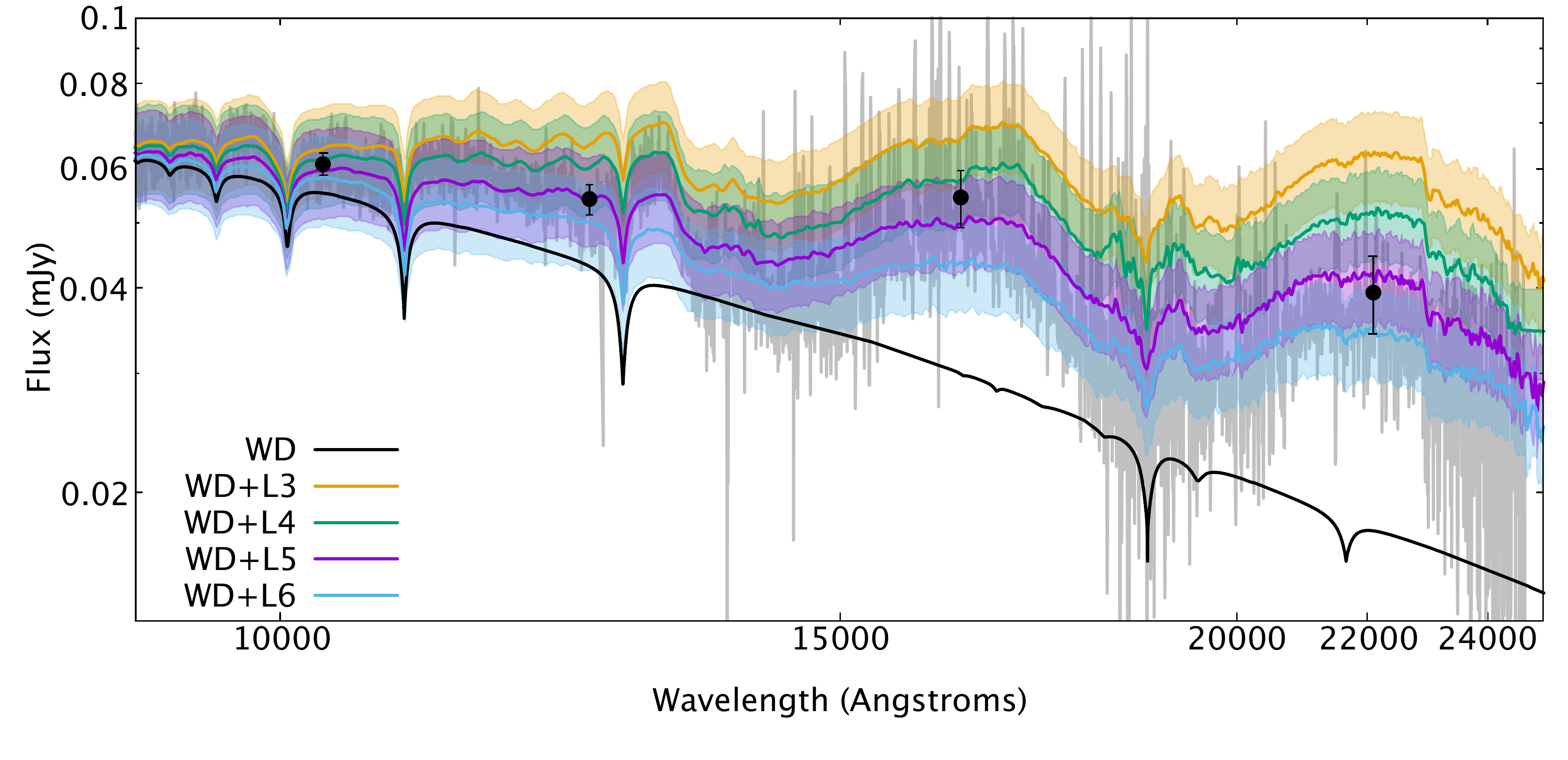}
    \caption{The GNIRS spectrum (grey) shown with the model white dwarf (black), and model white dwarf$-$brown dwarf spectra (orange, green, purple, blue). The UKIDSS photometry are also displayed. The shaded, coloured regions represent the 30 per cent rms scatter on the spectral types derived from the absolute magnitudes in \citet{dupuy12}.}
     \label{fig:nir}
     \end{center}
 \end{figure}

The GNIRS spectrum, templates and UKIDSS magnitudes are consistent with the L4-L6 spectral type suggested by \citet{steele11} once the 30 per cent rms scatter in the absolute magnitudes from \citet{dupuy12} is taken into account. Overall, a brown dwarf with a spectral type of  L5 is the most likely companion, although spectral types of L4 and L6 cannot be completely ruled out (Figure \ref{fig:nir}). 

%However, while all the models agree with an L5 spectral type, the $H$ band is more consistent with an L4-L5, and the $K$ band with an L5-L6. This is inconsistent with the results of \citet{casewell18} and \citet{casewell15} which show a possibly UV-induced $K$ band brightening in the irradiated brown dwarfs 
%SDSS~J141126.20$+$200911.1 and WD0137-349B.  It may be that the UKIDSS $K$ band photometry was obtained during part of the eclipse or that there is a reflection effect in the system.   There is, however, little or no reflection effect seen in the optical lightcurves suggesting the NIR magnitudes are unlikely to change by much over the whole orbit. Indeed, using our \textsc{lcurve} model to generate a model in the NIR, the reflection effect is only predicted to be * in the $JHK$ bands.

%\begin{figure}
 %\begin{center}
% \includegraphics[width=\columnwidth]{all_SED.pdf}
    %\caption{Infrared spectrum of WD1032+0l1 shown with white dwarf+brown dwarf %template spectra. The SDSS and UKIDSS photometry are also displayed.}
   %  \label{fig:sed}
  %   \end{center}
 %\end{figure}

\subsection{Masses and radii}

 \begin{figure*}
 \begin{center}
 \includegraphics[scale=0.7, trim = 2cm 0cm 0cm 0cm, clip]{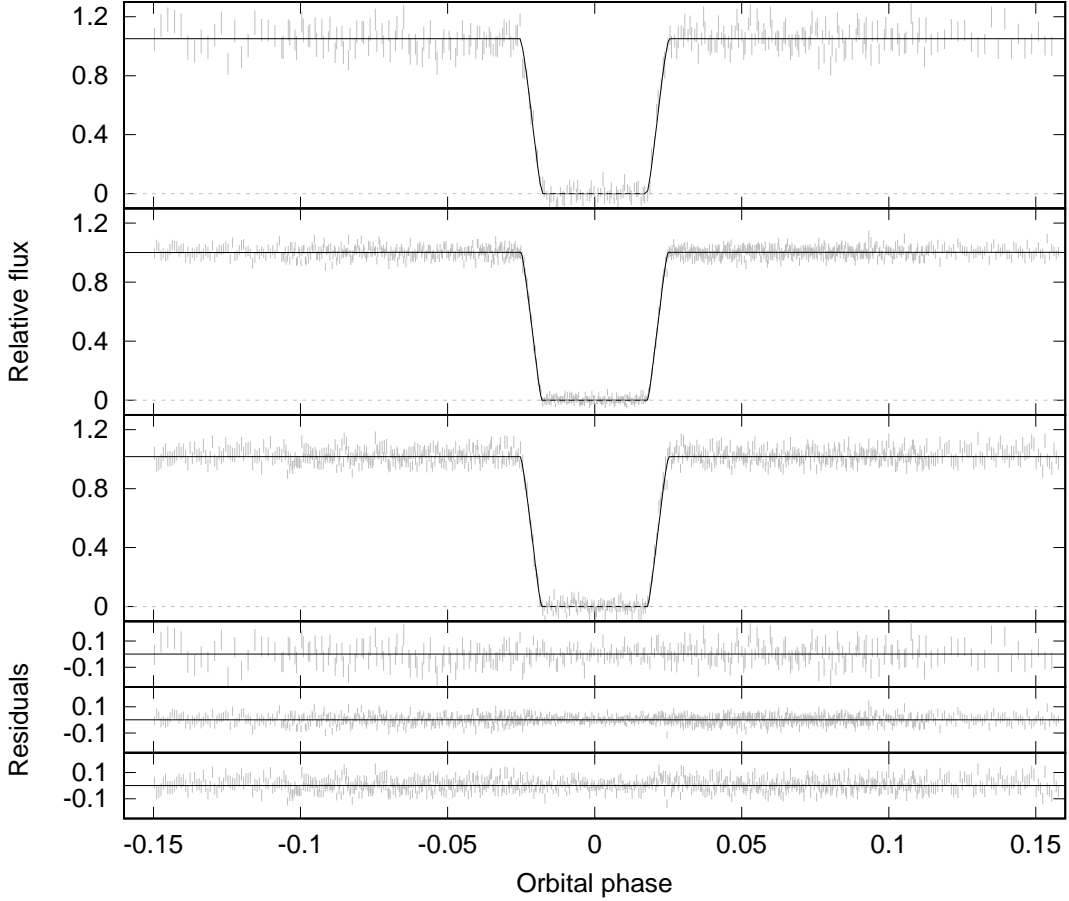}
    \caption{$u'$ (top), $g'$ (middle) $r'$ (bottom) lightcurves from all three nights of ULTRACAM data, with the best fitting model and residuals in each band. The out of eclipse flux was normalised to 1.}
     \label{fig:ugr}
     \end{center}
 \end{figure*}
 
 We normalised the ULTRACAM lightcurves to have a continuum level of 1, and phase folded using the ephemeris derived above in section 3.1.  We then fitted the ULTRACAM lightcurves using the method described in \citet{littlefair14} using an affine-invariant MCMC sampler (\textsc{emcee}; \citealt{foreman13}) and the lightcurve fitting code  \textsc{lcurve} \citep{copperwheat10}.  We used 100 walkers, with a burn-in period of 300, and 300 production steps. We used our parameters of the white dwarf and the  \citet{steele11} parameters for the brown dwarf to find the quadratic limb darkening coefficients with $T_{eff}$=9950~K, log $g$=7.65 from \citet{gianninas} for the white dwarf  and  $T_{eff}$=1700~K, log $g$=5.0 from \citet{claret12} for the brown dwarf. We only used the ULTRACAM data for this fitting as it has the highest cadence and smallest scatter which is needed to properly fit the ingress and egress of the eclipses. 

We allowed the mass ratio ($q$), radii ($R_1/a$, $R_2/a$), inclination ($i$) and the quadratic limb darkening parameters to vary. We put priors on the limb darkening parameters using a Gaussian distribution with twice the standard deviation suggested from the SED fit and the limb darkening tables relevant to each star. A half-Gaussian prior was also used for the mass ratio, conservatively set with a mean of 0 and a standard deviation of 0.3 using the white dwarf mass derived from the SED and a brown dwarf mass from the \citet{baraffe03} models for spectral type L5 (see previous section). Uniform priors were used for the radii and inclination. The acceptance fractions were 53 per cent for the $g$ band, 51 per cent in the $u$ band and 51 per cent in the $r$ band.

From the mass of the white dwarf, the orbital period and the radial velocity of the white dwarf we construct the mass function, and combine with the inclination from the lightcurve fit to find the brown dwarf mass and the orbital separation, which follows from Kepler's law. The separation is used to rescale the radii from the lightcurve fits to give the final parameters of the system which can be found in Table \ref{tab:res}, with the final fits shown in Figure \ref{fig:ugr}. The corner plots for the fits are given in Appendix B.  The mass and radius we derive for the white dwarf are consistent with both He and CO-core white dwarf models, but agree best with a CO model 
with a thin (10-10 M$_{\odot}$) envelope.

\begin{table*}
	\centering
	\caption{Final system parameters for WD1032$+$011.}
	\label{tab:res}
	\begin{tabular}{lcc} % 3 columns, alignment for each
    \hline
	Parameter	&	Value& Info\\
	\hline
	%this has been updated
WD $T_{eff}$(K)&9950 $\pm$ 150& SED\\
WD log  $g$&7.65 $\pm$ 0.13 &SED\\
$E(B-V)$&0.03 $\pm$ 0.01&SED\\
Cooling time (Gyr)&0.455 $\pm$ 0.080& FORS\\
$T0$ (BMJD)& 58381.2439008(10)  & ULTRACAM,ProEM \\
%P (days) &0.0915590004(3)& K2\\
$P$ (days)&0.09155899610(45)  &ULTRACAM,ProEM \\
$\gamma_1$ (kms$^{-1}$)&122.08 $\pm$ 1.94~ &GMOS\\
$K_1$ (kms$^{-1}$) &48.8 $\pm$ 2.64& GMOS\\
%q &0.09713$\pm$0.011&ULTRACAM\\
inclination($^\circ$)& 87.5 $\pm$ 1.4& ULTRACAM\\
$R_1$ (R$_\odot$)&0.0147 $\pm$ 0.0013& ULTRACAM\\
$R_2$(R$_\odot$)&0.1052 $\pm$ 0.0101& ULTRACAM\\
$M_1$(M$_\odot$)&0.4502 $\pm$ 0.0500& SED\\
$M_2$(M$_\odot$)&0.0665 $\pm$ 0.0061   &ULTRACAM \\
$a$(R$_\odot$)&0.6854 $\pm$ 0.0244&ULTRACAM\\
%$74.21\pm$5.58
\hline
\end{tabular}
\end{table*}

The constraint on the inclination from a single eclipse lightcurve is subtle. The ingress and egress duration depend upon the inclination, but are highly degenerate with the radii of the two components. Small changes in the shape of the ingress and egress can break this degeneracy, but the shape is also dependent on the limb-darkening of the white dwarf. This can make convergence difficult, and raise doubts about the constraint on the inclination from a primary eclipse alone. We checked against convergence issues by re-fitting the primary eclipse using a different parameterisation. We allowed the mid-eclipse time T0 to be a free parameter and parameterised the eclipse using the mass of the white dwarf, temperature of the white dwarf, the radius of the white dwarf and the radius ratio of the binary.  At each step in the MCMC chain we calculate the relevant limb darkening coefficients from the white dwarf temperature and the tables of \citet{gianninas}. We also checked our convergence by running the fits from different starting points, and by using both a quadratic and a Claret limb darkening law. All fits were consistent and gave posterior estimates of the brown dwarf mass and radius of  $M_2$= 0.066 $\pm$ 0.005 M$_{\odot}$ and 
$R_2$ = 0.099 $\pm$ 0.003 R$_{\odot}$, which is consistent with our original determination. The only main difference is that this parameterisation cannot rule out an inclination of 90 degrees for the binary, which is probably due to the relaxed constraint on $T0$.

%YJ observed together 2009-12-16 14:11:45.1 Y and 2009-12-16 14:36:14.1	J
% HK 2008-03-24 10:28:11.1 H and 2008-03-24 10:50:32.3 K

\section{Discussion}
In order to constrain the age of WD1032$+$011, we have performed a kinematic analysis using the proper motions from $Gaia$ and our measured velocities.  The cooling age of the white dwarf from the \citet{panei07} models is 800~Myr, so this gives us the minimum age of the system. We calculate the UVW space motions with respect to the local standard of rest to be,  U = $-$163$\pm$19 km\,s$^{-1}$, V = $-$73$\pm$2 km\,s$^{-1}$ and  W = 37 $\pm$ 8 km\,s$^{-1}$ (where U is positive towards the Galactic centre).  Using the same method we used in \citet{littlefair14} and the membership probabilities in \citet{bensby14} for memberships of the thin disk, thick disk and halo, we determine that WD1032$+$011 is 130 times more likely to belong to the thick disk than the halo, and 20000 times more likely to belong to the thick disk than the thin disk. Hence WD1032$+$011 is likely to belong to the thick disk but we cannot entirely rule out halo membership. This result means that the system is probably old, with a likely age of $\sim$10~Gyr if a member of the thick disk \citep{kilic19}. \citet{gallart19} shows that for the  kinematics of WD1032$+$011 and any radial velocity between $-$120 to $+$120 km\,s$^{-1}$ ($\gamma$=122.08 $\pm$ 1.94 km\,s$^{-1}$) thick disk membership is favoured, strongly suggesting an age greater than 5~Gyr and a moderately low metallicity of [Fe/H]$\sim-0.3$. 

With the 5$-$10~Gyr age estimate from the kinematic analysis we can compare our brown dwarf mass and radius to the low metallicity ([Fe/H]=-0.5) Sonora-Bobcat models. Using these models a 10~Gyr-old  0.066 M$_{\odot}$ brown dwarf would have $T_{eff}$=988~K and a radius of 0.079~R$_{\odot}$. Similarly, a 6~Gyr-old 0.066~M$_{\odot}$ brown dwarf would have $T_{eff}$=1163~K and a radius of 0.079~R$_{\odot}$. These models suggest that WD1032$+$011 should have a much smaller radius than the one we measure here (0.1052 $\pm$ 0.0101 $R_{\odot}$).  The effective temperatures are also much lower than one would expect for an L5 dwarf.  A 1500 K (approximate effective temperature of an L5 dwarf) 0.066~M$_{\odot}$ brown dwarf would have an age of only 2 Gyr according to the Sonora-Bobcat models, but should also have a radius of 0.0851~R$_{\odot}$, again smaller than our measured radius. It is possible that the system has an age of 2~Gyr, but this would be unusually young for a thick disk member. We therefore conclude that WD1032$+$011 is likely hotter and larger than the models predict, making it the first inflated brown dwarf to be discovered orbiting a white dwarf.

\begin{figure*}
 \begin{center}
 \includegraphics[scale=0.35]{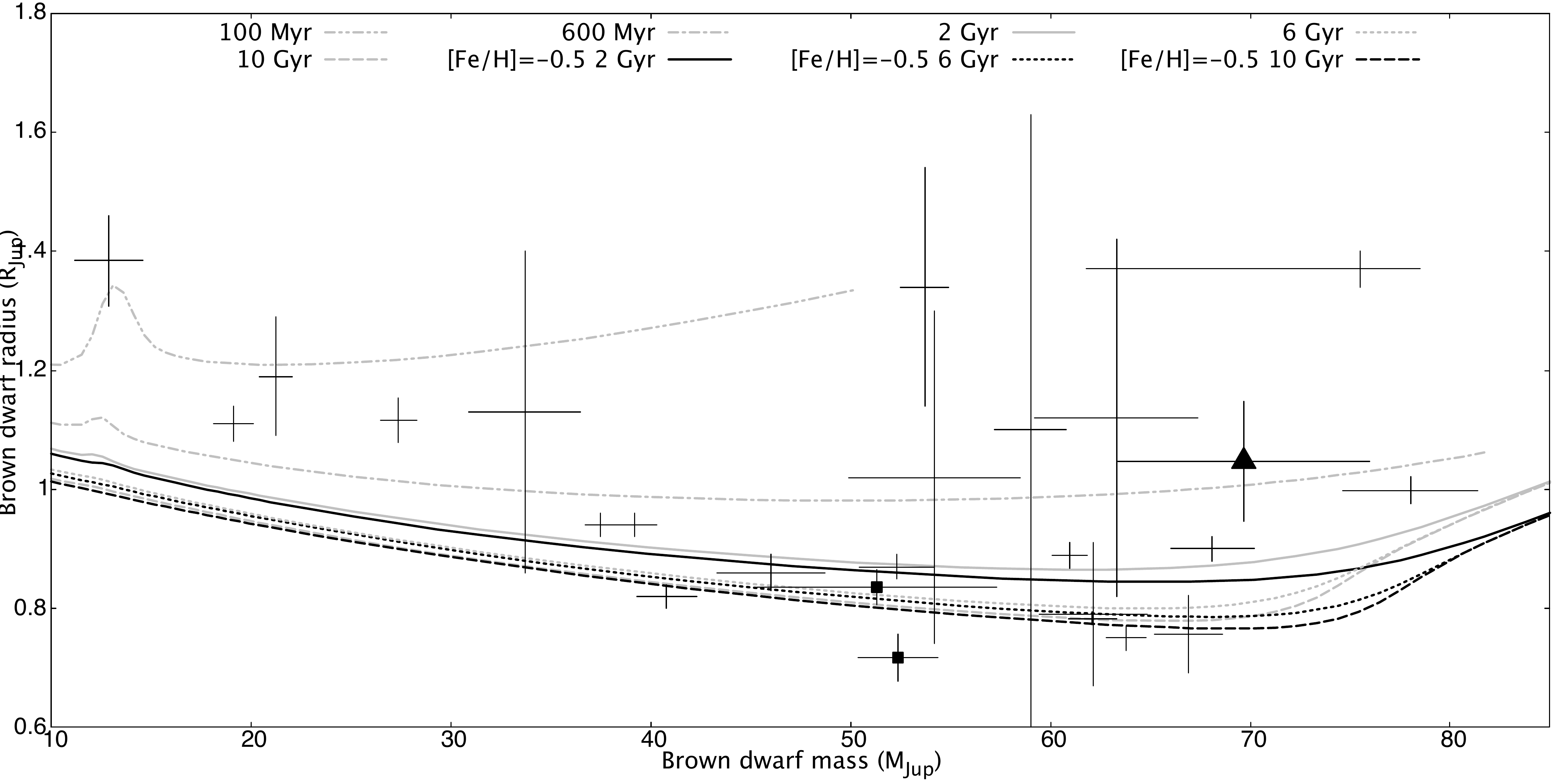}
    \caption{Masses and radii for all brown dwarfs transiting white dwarfs (filled squares; \citealt{littlefair14, parsons17}), and main sequence stars (\citealt{carmichael20} and references therein). WD1032 is marked with a filled triangle. Also shown are the Sonora-Bobcat evolutionary models of \citet{sonora_web} for 100 Myr, 600 Myr, 2 Gyr, 6 Gyr and 10 Gyr with solar metallicity (grey) and low metallicity ([Fe/H]=$-$0.5) models for 2 Gyr, 6 Gyr and 10 Gyr (black).}
     \label{MR}
     \end{center}
 \end{figure*}

WD1032$+$011 is only the third white dwarf-brown dwarf binary where the radius of the brown dwarf can be directly measured (Table \ref{bin}).   Both the previously known eclipsing brown dwarfs,  SDSS~J141126.20$+$200911.1 and SDSS~J120515.80$-$024222.6, show no inflation and are consistent with the 6-10~Gyr Sonora-Bobcat isochrones from \citet{sonora_web} (Figure \ref{MR}). WD1032$+$011 is the first brown dwarf in a white dwarf-brown dwarf binary to have been shown to be  inflated, which is extremely interesting as the mechanism causing the inflation is unknown. SDSS~J120515.80$−$024222.6 has a much hotter white dwarf primary, and a much shorter orbital separation than WD1032$+$011 yet is not inflated, indicating that any inflation cannot be due to irradiation alone. We also do not see any signs of interaction between the white dwarf and brown dwarf in WD1032$+$011 as is seen for NLTT5306AB \citep{longstaff19} where the white dwarf shows emission features due to weak accretion from the brown dwarf, possibly due to a wind.

There are also four brown dwarfs known to be eclipsing hot sdB stars \citep{geier11, schaffenroth14, schaffenroth15}. It is, however, challenging to determine the mass and radius of hot subdwarfs as they are often pulsating and there is no well defined mass-radius relationship as there is for white dwarfs.  Large uncertainties regarding the mass and radius of the primary can cause large errors on measurements of the brown dwarf, meaning radii from mass-radius relations are often adopted. For this reason we do not discuss brown dwarfs in binaries with hot subdwarfs further here. There are however $\sim$20 systems where a brown dwarf eclipses a main sequence star. These systems have been discovered through transiting planet searches, and do have reliable masses for the primary stars.  The mass-radius relationship for all 23 transiting, irradiated brown dwarfs is shown in Figure \ref{MR}.

When we compare the three brown dwarfs orbiting white dwarfs (filled boxes and circle in Figure \ref{MR}) to the population of irradiated brown dwarfs orbiting main sequence stars  it is clear that most, if not all, of the low mass objects ($M\lesssim35$ M$_{\rm Jup}$) are inflated. None of these brown dwarfs orbit a star that has been identified as younger than 1 Gyr, as the radii of the brown dwarf would suggest.  At masses greater than 35 M$_{\rm Jup}$, the majority of objects sit on the 5-10 Gyr isochrone. The exceptions are:  NGTS-7Ab (\citealt{jackman19}; $M=75.5^{+3}_{-13.7}$ M$_{\rm Jup}$) which is $\sim$ 55 Myr old, hence its position near, but below the 100 Myr isochrone;  TOI0-503 (\citealt{subjak19}; $M=53.7\pm1.2$ M$_{\rm Jup}$) which is 180 Myr old and has a radius consistent with this; KOI-189b, which may in fact be a low mass star (\citealt{diaz14}; $M=78\pm3.4$ M$_{\rm Jup}$), and may be slightly inflated, as age estimates for this system are $\sim$5~Gyr. However, inflated late M-dwarf radii are not uncommon, and are often attributed to convection within the star being inhibited due to magnetic fields (e.g. \citealt{macdonald14}).  The remaining two objects that do not sit on the 5--10 Gyr isochrones are  CoRoT-33b \citep{cszimadia} and CoRot-15b \citep{bouchy}. Both of these objects have large uncertainties on their radii, but also orbit active stars which may have had some effect on the measurement of the radius of the brown dwarf.

\citet{parsons18} found that the scatter in the M-dwarf Mass-Radius relationship was 6.2$\pm$4.8 per cent, with only about a quarter of M dwarfs being consistent with models. They determined that there was no trend with either age or metallicity as to which M dwarfs are inflated. It may be that a similar relationship, with similar scatter exists as we move into the brown dwarf regime, particularly for the higher mass brown dwarfs.

\begin{table*}
	\centering
	\caption{Eclipsing white dwarf-brown dwarf binaries. The system parameters are from \citet{littlefair14,beuermann13,parsons17} and this work.}
	\label{bin}
	\begin{tabular}{lccccccc} % 3 columns, alignment for each
    \hline
Name& Period& $M_{1}$& $R_{1}$& $T_{eff}$& $M_{2}$& $R_{2}$& spectral type\\
&hr&$M_{\odot}$&$R_{\odot}$&K&$M_{\odot}$&$R_{\odot}$&\\
\hline
WD1032$+$011&2.20&0.450$\pm$0.050&0.0148$\pm$0.0013&9950 $\pm$150&0.067$\pm$0.006&0.105 $\pm$ 0.010&L5\\
SDSS~J141126.20$+$200911.1&2.03&0.53$\pm$0.03&0.0142$\pm$0.0006&13000$\pm$300&0.050$\pm$0.002& 0.072$\pm$0.004&T5\\
SDSS~J120515.80$−$024222.6 &1.19&0.39$\pm$0.02&0.0217--0.0223&23680$\pm$430&0.049$\pm$0.006&0.081--0.087&$>$L0\\
\hline
\end{tabular}
\end{table*}
%WD Mass = 0.450273 +/- 0.050047
%WD Rad  = 0.014759 +/- 0.001325
%BD Mass = 0.066541 +/- 0.006146
%BD Rad  = 0.105197 +/- 0.010108

\section{Conclusions}
We have discovered a new eclipsing, detached short period white dwarf-brown dwarf binary member of the thick disk. Our multi-colour lightcurves of the eclipses show that the brown dwarf is inflated when compared to metal poor evolutionary models. A Gemini GNIRS near-IR spectrum of the brown dwarf is consistent with a spectral type of L5 which would suggest an effective temperature hotter than predicted by the models for the age of the thick disk.

\section*{Acknowledgements}

We thank the GMOS team at Gemini-North, in particular Sandy Leggett, Siyi Xu and Andr\'e-Nicolas Chen\'e for their assistance and advice with the scheduling and data reduction. 

S.L. Casewell and S.G. Parsons acknowledge support from STFC Ernest Rutherford Fellowships and I.P Braker acknowledges support from the University of Leicester College of Science and Engineering. S.G. Parsons also acknowledges the support of the Leverhulme Trust. V.S. Dhillon and ULTRACAM are supported by the STFC grant SST/R000964/1. Partial support for this work was also provided by NASA K2 Cycle 6 grant 80NSSC19K0162. ZV acknowledges support from the Wooten Center for Astrophysical Plasma Properties (WCAPP) under DOE grant DE-FOA-0001634, and DW and KW acknowledge support from NSF Grant 1707419.

This work is based on observations obtained at the Gemini Observatory, which is operated by the Association of Universities for Research in Astronomy, Inc., under a cooperative agreement with the NSF on behalf of the Gemini partnership: the National Science Foundation (United States), National Research Council (Canada), CONICYT (Chile), Ministerio de Ciencia, Tecnolog\'{i}a e Innovaci\'{o}n Productiva (Argentina), Minist\'{e}rio da Ci\^{e}ncia, Tecnologia e Inova\c{c}\~{a}o (Brazil), and Korea Astronomy and Space Science Institute (Republic of Korea). Also based on observations collected at the European Organisation for Astronomical Research in the Southern Hemisphere under ESO programme 098.D-0717(A). This research is based on observations made with the Galex mission, obtained from the MAST data archive at the Space Telescope Science Institute, which is operated by the Association of Universities for Research in Astronomy, Inc., under NASA contract NAS 5–26555.
%%%%%%%%%%%%%%%%%%%%%%%%%%%%%%%%%%%%%%%%%%%%%%%%%%



\bibliographystyle{mnras}
\bibliography{bib} % if your bibtex file is called example.bib

\appendix
\section{Eclipse times}

\begin{table*}
	\centering
	\caption{Eclipse times of WD1032$+$011}
	\label{tab:eclipse}
	\begin{tabular}{cccccc}
		\hline
	    Date& BMJD(TDB) &Eclipse number& O-C (s)& Instrument\\
		\hline
		2017 Dec 18&58105.468197(16)&-3012&-0.6&ProEM\\
		2017 Dec 19&58106.4753509(54)&-3001&-0.2&ProEM\\
		2017 Dec 20&58107.390991(32)&-2991&4.0&ProEM\\
		2017 Dec 20&58107.482501(15)&-2990&-0.2&ProEM\\
		2017 Dec 21&58108.398083(14)&-2980&-0.9&ProEM\\
		2017 Dec 21&58108.489669(21)&-2979&1.5&ProEM\\
		2018 Jan 19&58137.2391769(33)&-2665&0.06&ULTRACAM\\
        2018 Jan 23&58141.1762135(35) &-2622&0.03&ULTRACAM\\
        2019 Mar 01&58543.3033239(17)&1770&-0.02&ULTRACAM\\
		\hline
		\end{tabular}
\end{table*}
\section{Cornerplots}
\begin{figure}
 \begin{center}
 \includegraphics[width=\columnwidth]{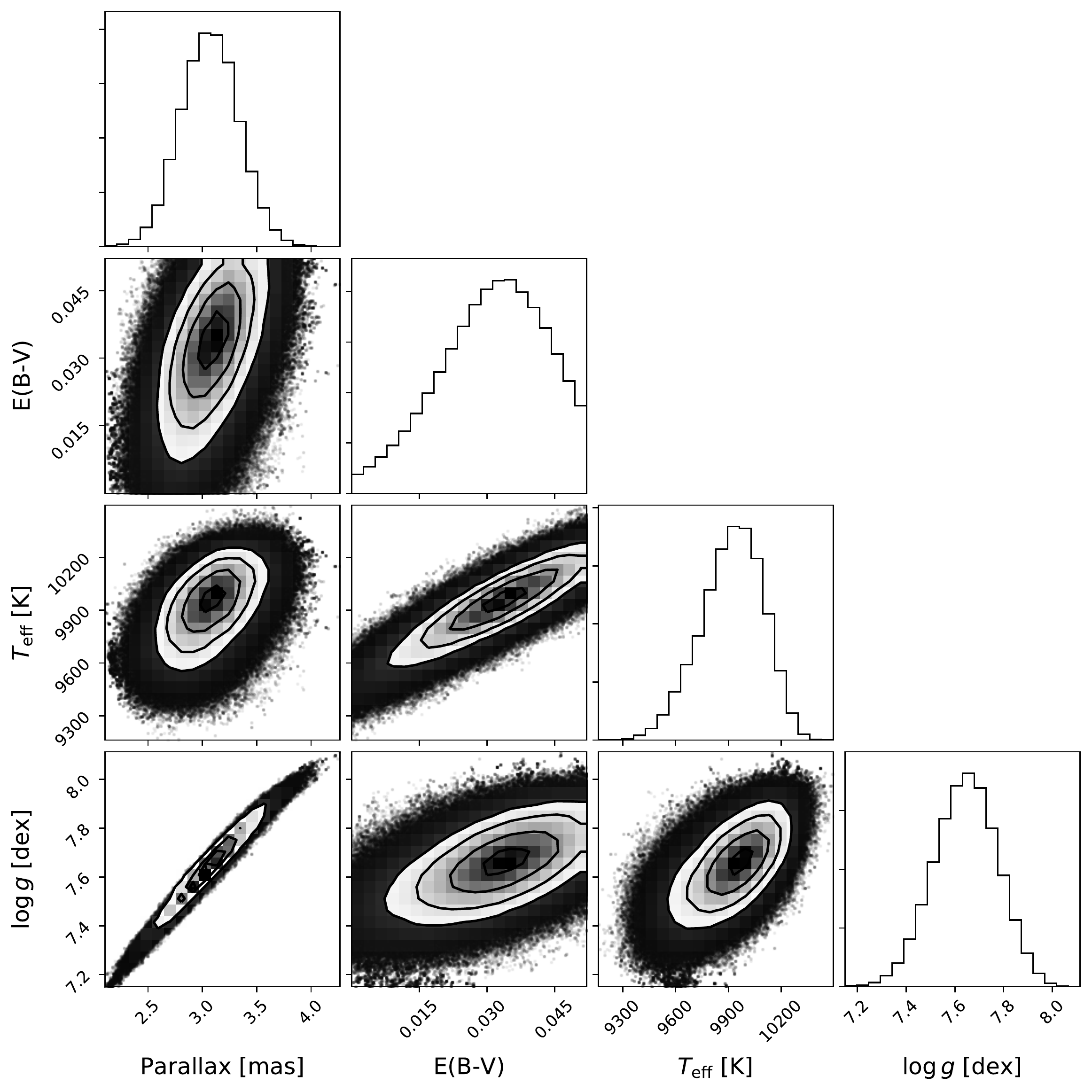}
    \caption{Corner plot from the MCMC output used to fit the SED of the white dwarf}
     \end{center}
 \end{figure}
 \begin{figure}
 \begin{center}
 \includegraphics[width=\columnwidth]{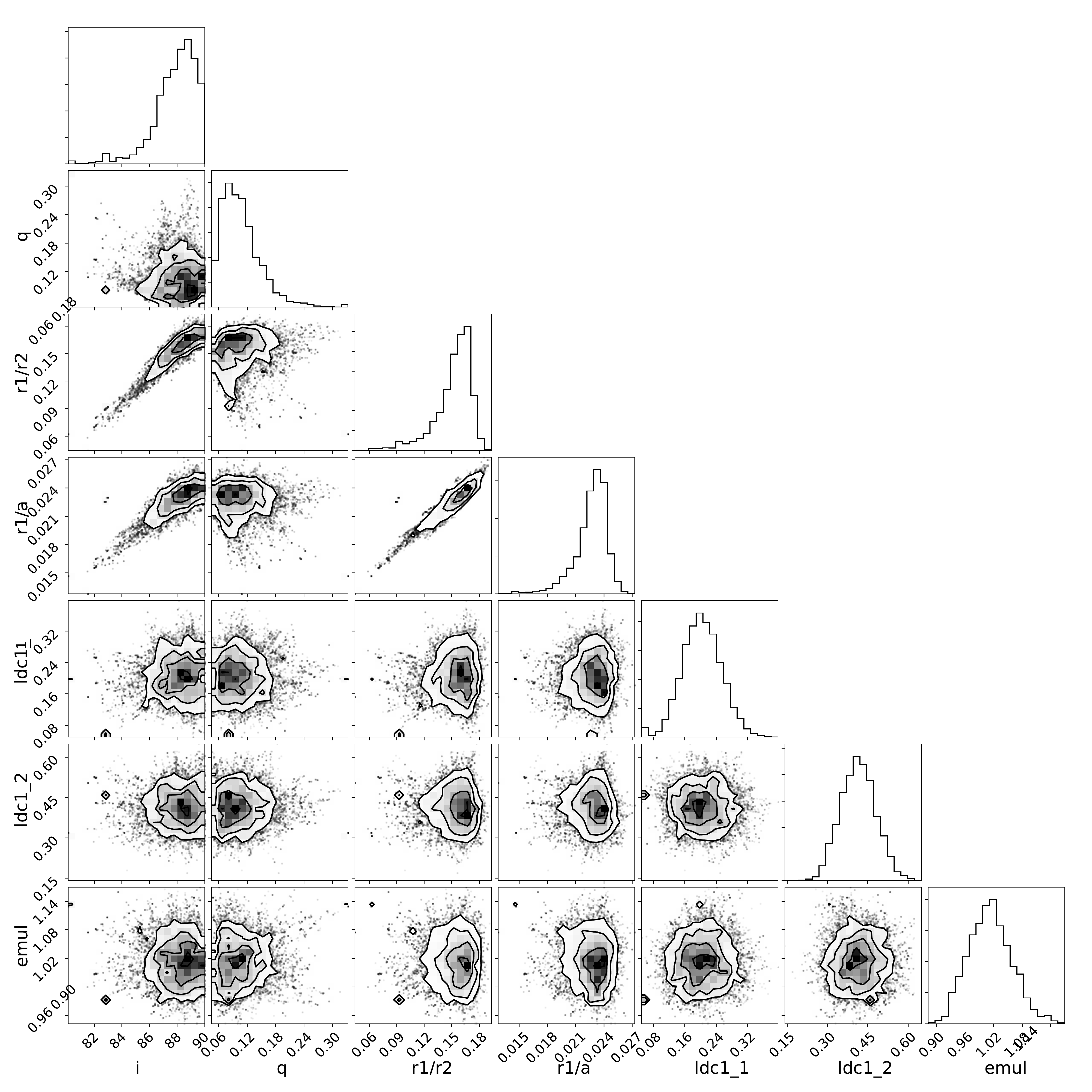}
    \caption{Corner plot from the MCMC output used to fit the $u$ band lightcurve}
     \end{center}
 \end{figure}

  \begin{figure}
 \begin{center}
 \includegraphics[width=\columnwidth]{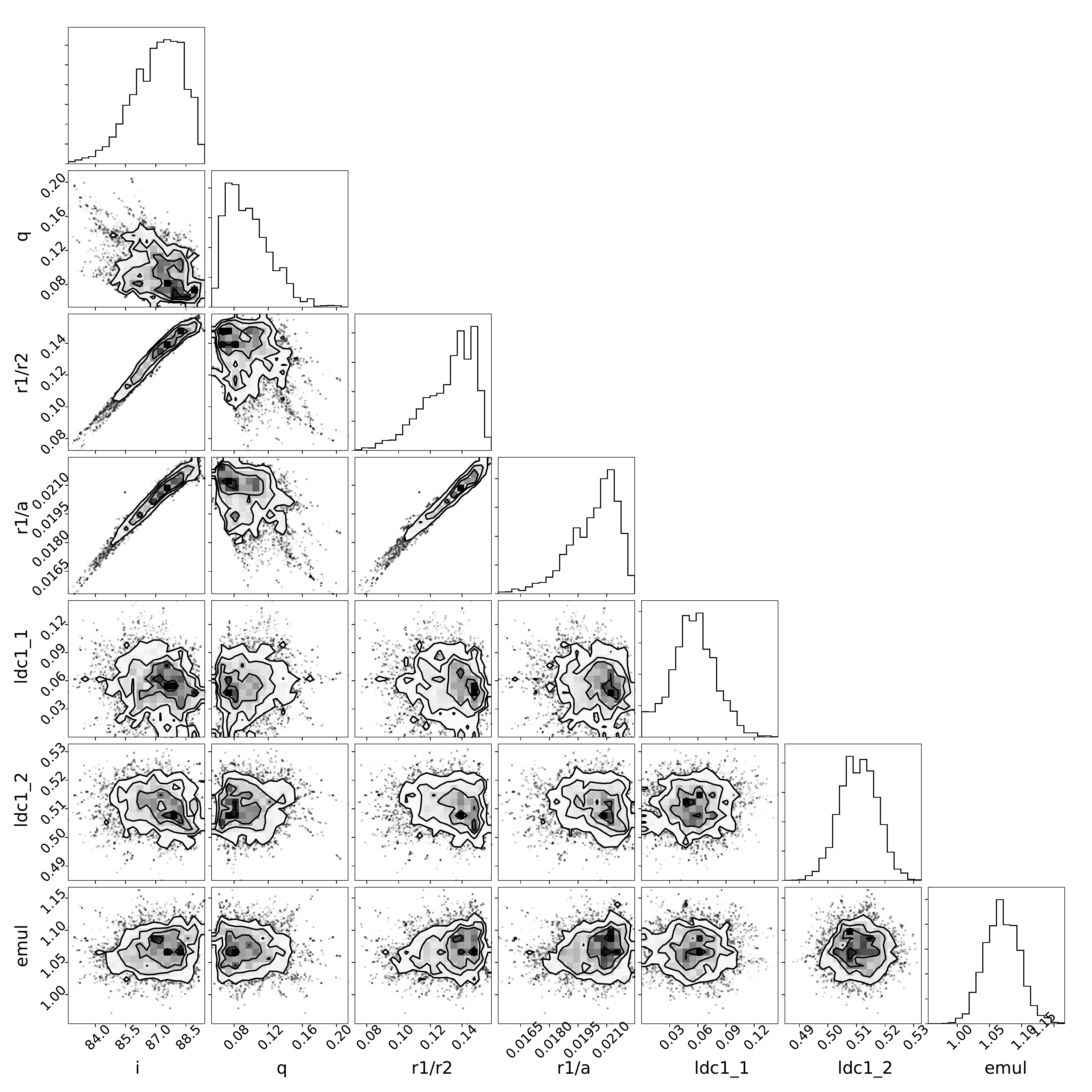}
    \caption{Corner plot from the MCMC output used to fit the $g$ band lightcurve}
     \end{center}
 \end{figure}

  \begin{figure}
 \begin{center}
 \includegraphics[width=\columnwidth]{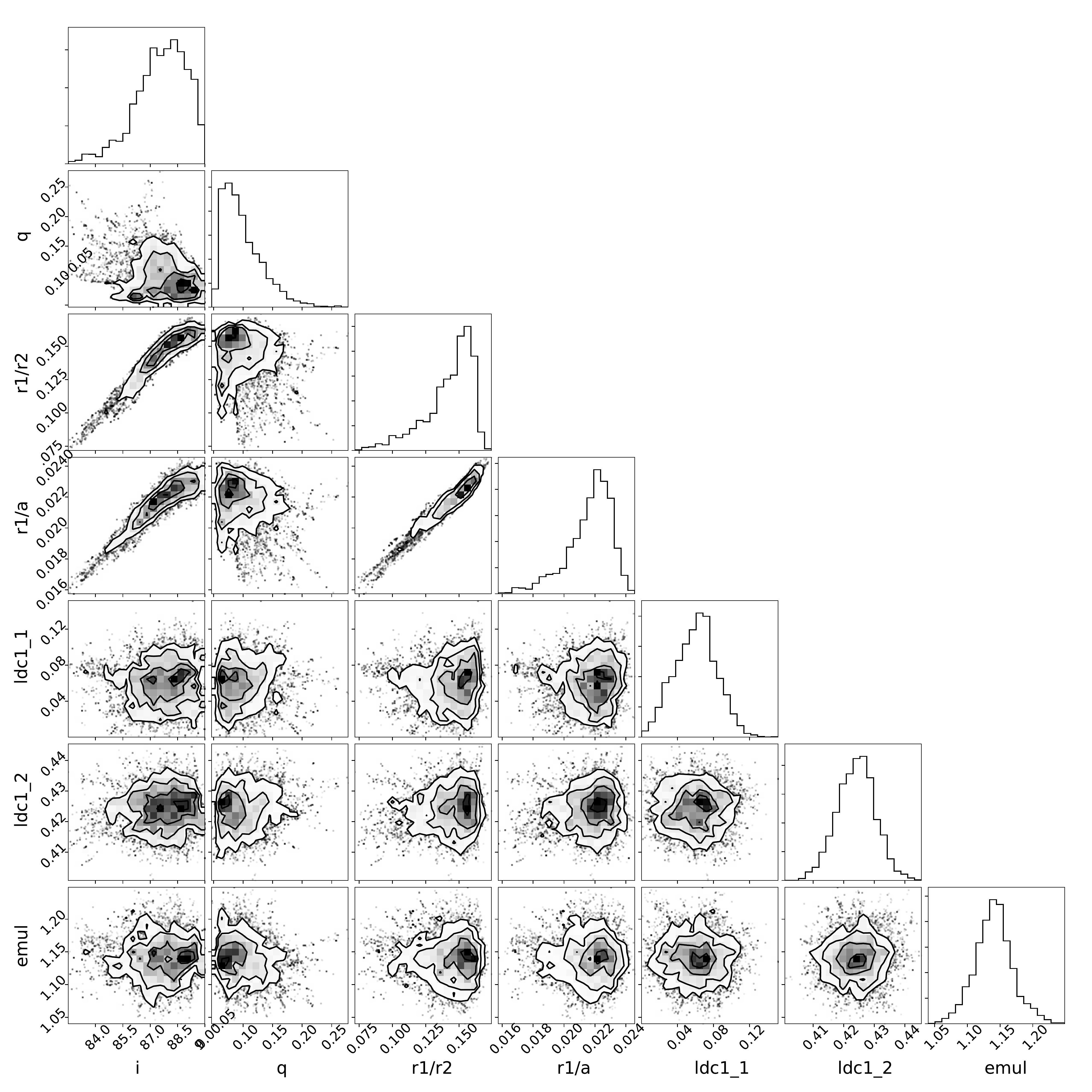}
    \caption{Corner plot from the MCMC output used to fit the $r$ band lightcurve}
     \end{center}
 \end{figure}

\bsp	% typesetting comment
\label{lastpage}
\end{document}